# On the Spin Gap Phase of Strongly Correlated Fermions

P. Lederer[1] and Elihu Abrahams[1,2]

[1] Physique des Solides, Université Paris-Sud, F91405 Orsay-cedex, France
Laboratoire associé au CNRS
[2] Serin Physics Laboratory, Rutgers University, P.O. Box 849
Piscataway, New Jersey 08855

We discuss the possible existence of a "spin-gap" phase in the low-doping regime of strongly-correlated two-dimensional electrons within the gauge field description of the $t-J$ model. The spin-gap phase was recently shown by Übbens and Lee to be destroyed by gauge field quantum fluctuations for a single-layer 2D system in the absence of disorder and for a full gap. We show that the same conclusion applies both in the dirty limit and for the case of a gapless spinon condensate.

PACS Nos. 74.30E; 64.70D

## I. INTRODUCTION

Strong electronic correlations have been described as essential to the understanding of the quasi-two-dimensional high-$T_c$ superconductors.[1] Spin-charge separation is a possible manifestation of strong interactions which has been proposed as basic to the description of the low-energy excitations in the 2D interacting electron system.[2] For strong interaction, the $t-J$ model[3] version of the Hubbard model is often used as a starting point. The system is studied when it is doped with $x$ holes per site away from half-filling (thus $1-x$ electrons per site). Configurations in this model are constrained to exclude those where the number of electrons on a given site can exceed unity. The slave boson method[4] is a way to deal with this strong-correlation constraint and it leads naturally to spin-charge separation via the introduction of $x$ charged bosonic holons and $1-x$ chargeless fermionic spinons. A mean-field approach yields the phase diagram sketched in Fig. 1. It contains a number of phases including a "strange metal phase" and phases in which either the spinons or the holons (or both) are condensed.[5] It has been suggested[6] that the properties of the normal state of high-$T_c$ superconductors can be well-described by the strange metal phase which is essentially a uniform RVB state.[1].

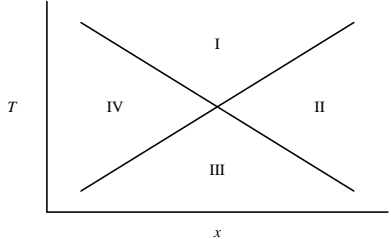

FIG. 1. Mean field phase diagram of the $t-J$ model in the temperature ($T$)-doping ($x$) plane, containing the "strange" metal phase (I), Fermi liquid phase (II), superconductor phase (III), and spin gap phase (IV).

A convenient way to discuss the phases is to consider the Ioffe-Larkin expression[7] for the response function $\Pi$ of the total system to an external electromagnetic field:

$$\Pi(q,\omega) = \frac{\Pi_f(q,\omega)\Pi_b(q,\omega)}{\Pi_f(q,\omega)+\Pi_b(q,\omega)}, \qquad (1.1)$$

where $\Pi_f$ and $\Pi_b$ are the current-current response functions of the fermion (spinon) and boson (holon) subsystems. This equation is a consequence of the strong-correlation constraints. The response of the total system is dominated by the smaller of $\Pi_f$ and $\Pi_b$. Accordingly, the schematic phase diagram shown in Fig. 1 has four parts. When both fluids are normal, i.e. when the spinons are a normal Fermi gas and the holons are not Bose-condensed, as initially studied in Ref. 4, we have a strange metal, with a Fermi surface for $1-x$ electrons which obeys Luttinger's theorem, but with transport properties determined by the doping $x$. Thus there is a positive Hall constant and a resistivity proportional to $1/x$. The temperature dependence of the resistivity, $\rho \propto T^\beta$ with $\beta$ near unity, is determined by the neglect of the holon diamagnetic susceptibility in comparison with the spinon diamagnetic susceptibility which enters the denominator of the $\Pi$ response functions.

The mean-field theory of the system accounts for the local constraints only globally. The corrections involve the fluctuations of a gauge-field which mediates the conservation laws of the bosons and fermions which are implied by the constraint. The reason for the unusual resistivity law is that the less-conducting particles, the bosons, are inelastically scattered by low-frequency long-wavelength gauge fluctuations.

The retarded gauge field propagator $D(q,\omega)$ is given by:

$$D^{-1} = \Pi_f + \Pi_b$$
$$\Pi_{f,b} = \chi_{f,b} q^2 - i\omega \sigma_{f,b}(q,\omega). \tag{1.2}$$

The spinon (fermion) and holon (boson) susceptibilities are

$$\chi_f = \frac{1}{12\pi m_f}$$
$$\chi_b = \frac{n_b}{12 m_b^2 T}. \tag{1.3}$$

In the pure case, the spinon and holon conductivities are

$$\sigma_f(q,\omega) = \frac{2 n_f}{m_f q v_f}, \quad \omega < q v_f$$
$$\sigma_b(q,\omega) = \frac{2 n_b \sqrt{\pi}}{m_b q v_b}, \quad \omega < q v_b = q\sqrt{2T/m_b}, \tag{1.4}$$

where $v_b = \sqrt{2T/m_b}$ is the thermal velocity of bosons. In the above equations, the fermion and boson areal number densities and masses appear. They are given in terms of the energy parameters of the $t-J$ model and the doping $x$ away from half-filling as

$$(n_f/m_f) = (k_F^2/2\pi m_f) \simeq 2(1-x)J$$
$$(n_b/m_b) \simeq 2xt, \tag{1.5}$$

where $k_F$ is the Fermi momementum of spinons. The conductivities are zero for $\omega > q v_{f,b}$. In Eqs. (1.2), we have assumed that the temperature is sufficiently high that the bosons are far from condensation and we have treated their distribution function classically. For non-interacting bosons, this would require that $2xt/T < 1$. However, the interactions extend this behavior to a wider temperature range.[8]

In the strange metal phase, the fermionic contribution dominates the gauge fluctuations and we have

$$D^{-1} \simeq \Pi_f = 2(n_f/m_f)\left[-i(\omega/q v_f) + (q^2/12 k_f^2)\right] \tag{1.6}$$

Übbens and Lee[9] have recently argued that in the pure system the spin-gap phase (region IV in Fig. 1) is an artifact of the mean field approximation. According to them, the spin-gap is destroyed by the quantum fluctuations of the gauge field. Their argument goes as follows: The possible condensation of the spinons into the spin-gap phase is like that of the superconducting transition. As in BCS theory, one can compute the negative contribution $F_{cond}(\Delta)$ of the condensation to the total free energy as a function of the spinon pairing amplitude $\Delta$. However, the free energy has a contribution from the gauge field as well[10] because when a spinon gap opens, a mass is induced in the fermionic contribution to the gauge fluctuation spectrum and the low-lying gauge field fluctuation contribution to the free energy is lost. This gives a *positive* contribution to the free energy. It is[9,10]

$$F_{gauge}(\Delta) - F_{gauge}(0) = \int \frac{d^2 q}{(2\pi)^2} \int_0^\infty d\omega \coth\left(\frac{\omega}{2T}\right) \arctan\left[\frac{\mathrm{Im} D^{-1}(q,\omega,\Delta)}{\mathrm{Re} D^{-1}(q,\omega,\Delta)}\right] - F_{gauge}(0). \tag{1.7}$$

It is necessary to examine the effect of $\Delta$ on $D^{-1} = \Pi_f + \Pi_b$. In lowest order, only $\Pi_f$ is affected. As in superconductivity, the massless $\Pi_f$ of the normal phase [see Eq. (1.2)] acquires a gap proportional to the superfluid density (Meissner effect). We have $\Pi_f(\Delta) = n_{sf}/m_f + O(\omega, q^2)$. Here $n_{sf} = n_f$ at $T=0$ in the clean (London) limit and $n_{sf} = n_f \Delta \tau_f$ in the dirty limit ($\Delta \tau_f < 1$), where $\tau_f$ is the spinon elastic transport lifetime (see Ref. 11, for example). In fact, $\mathrm{Im}\Pi_f(q,\omega)$ and $\mathrm{Re}\Pi_f(q,\omega)$ become complicated functions of $\omega$, especially when the gap function is anisotropic. The main effect on the low frequency part of $\Pi_f$ is that $\mathrm{Im}\Pi_f = 0$ for $\omega < 2\Delta$ at $T=0$. With neglect of possible spinon gap anisotropy, an estimate of $F_{gauge}(\Delta) - F_{gauge}(0)$ in the clean limit is given by the contribution to $F_{gauge}(0)$ of the gauge-field modes with $\omega < 2\Delta$ which are missing in the condensed phase[9]:

$$F_{gauge}(\Delta) - F_{gauge}(0) \simeq \int_0^{2\Delta} d\omega \coth(\frac{\omega}{2T}) \int \frac{d^2q}{(2\pi)^2} \arctan\left(\frac{n_f \omega}{m_f \chi q^3}\right)$$

$$\simeq \int_0^{2\Delta} d\omega \coth(\frac{\omega}{2T}) \left(\frac{n_f \omega}{m_f \chi}\right)^{2/3}$$

$$\simeq \left(\frac{n_f}{m_f \chi}\right)^{2/3} \Delta^{5/3}, \tag{1.8}$$

where $\chi$ is the coefficient of $q^2$ in Re$D^{-1}$, $\chi = (1/12\pi m_f) + (n_b/12 m_b^2 T)$. Since $F_{gauge}(\Delta) \sim \Delta^{5/3}$ is non-analytic in $\Delta^2$, the spinon pairing transition must be actually be first order. More importantly, $F_{gauge}(\Delta) - F_{gauge}(0)$ is always much larger than $|F_{cond}(\Delta)|$ for any $\Delta$, as long as $\chi$ remains small. This implies that the opening of a gap does not happen for $T > T_{BE} = 2\pi n_b/m_b$. For completeness, we have reproduced here the argument of Übbens and Lee[9] but we included the holon contribution to Re$D^{-1}$ in Eq. (1.8). This reduces very slightly the impact of $F_{gauge}$ but does not change the conclusion.

One might wonder if the vanishing of the mean-field spin gap is a property of the pure system alone, since low-frequency gauge fluctuations are significantly modified in the dirty limit. The conductivities will be finite as $q \to 0$, contrary to the pure case. As a result, the zero-point fluctuations should be significantly reduced in the long wavelength limit. The most favorable situation for the preservation of a mean-field spin pseudogap would be a gapless situation, whereby a finite density of excited states continues to exist down to zero energy.

The purpose of this paper is to examine this question. Our conclusion is that the mean-field spin gap phase is also suppressed in the dirty limit of the spinon superfluid, as well as in the gapless case. Although there remain low frequency gauge fluctuations associated with the holon liquid, for $T > T_{BE}$, the suppression of the low frequency modes in the normal phase proves too costly, in comparison with the gain in spinon gap condensation energy, to allow the condensation to happen.

## II. DIRTY LIMIT

### A. Normal phase

In the normal phase, disorder produces a non-zero long-wavelength conductivity for each component of the plasma. In a Fermi liquid, electron-impurity scattering gives a conductivity $n_f \tau_e/m_f$ with an elastic scattering rate $\tau_e^{-1} = 2\pi n_i N_0 |V|^2$. Here, $V$ is the Fourier coefficient of the weak short range electron-impurity potential, $n_i$ is the density of impurities and $N_0$ is the one-spin density of states. The effect of impurity scattering in the case of the strongly correlated electron gas has been treated extensively by Ioffe and Kotliar[12] (IK) in the context of the $t-J$ model. There also exist discussions for the Hubbard model.[13] Following IK, we express the electron scattering rate in terms of the variance $w$ of the random potential: $\tau_e^{-1} = 2\pi N_0 w^2$. It was pointed out by IK that in the strongly-correlated electron gas, impurity scattering results in quantitatively different transport lifetimes for the fermions and the bosons. The interplay between the non-uniformity in space of the random potential and the strong-correlation constraints creates non-uniform chemical potentials (Lagrange multipliers) for the spinons and holons. The holons see an unscreened random potential, while the effective spinon-impurity potential is screened by the backflow of the holons; it becomes smooth and only forward scattering is appreciable. Consequently, the spinon transport lifetime is much longer than the boson one:

$$\tau_f \simeq \frac{24\pi t^2}{J w^2 x^{3/2}}, \tag{2.1}$$

while $\tau_b \simeq 2t/w^2$, so that $\tau_f/\tau_b \simeq 12\pi x^{-3/2}(t/J) >> 1$. As a result, as emphasized by IK, the resistivity is determined by the holons.

The fermion conductivity $\sigma_f = n_f \tau_f/m_f$ is determined by the above value of $\tau_f$ as long as $q v_f \tau_f < 1$. Otherwise, the pure value given in Eq. (1.4) obtains. In fact, the fermion conductivity departs from its value in the pure case only in a small region of reciprocal space, i.e. for $q/k_f < 10^{-2}$ to $10^{-3}$, if we choose $w \simeq t$ and $x \simeq .1 - .2$. On the contrary, the boson conductivity is governed by impurity scattering in the whole reciprocal space.

We conclude that the gauge field propagator $D(q,\omega) = \Pi_f(q,\omega) + \Pi_b(q,\omega)$ is practically determined entirely by the spinon polarisability, while the physical conductivity is dominated by the holon fluid:

$$\sigma = \frac{\sigma_f \sigma_b}{\sigma_f + \sigma_b} \simeq \sigma_b = 4x(t/w)^2. \tag{2.2}$$

Then the free energy cost due to a mean field spin gap $\Delta$ has the form originally given in Ref. 9 and quoted here in Eq. (1.8):

$$-F_n = 38.3 N(0)\Delta^2 \left[\frac{(1-x)J}{\Delta}\right]^{1/3}, \qquad (2.3)$$

where $N(0)$ is the one-spin density of states, $m_f/2\pi$.

### B. The spin gap phase

In the spin gap phase, in the dirty case, a free energy contribution from gauge fluctuations is recovered. the fermion polarisability is that of a dirty superfluid, i.e. $\text{Im}\Pi_f = 0$, $\text{Re}\Pi_f = (n_f \Delta \tau_f / m_f) + 0(\omega)$, valid if $\Delta \tau_f < 1$. Since the boson polarisability is unchanged compared to the normal phase as long as $T > T_{BE}$, the inverse gauge field propagator is now:

$$D(q,\omega)^{-1} \simeq (n_f/m_f)\Delta\tau_f - i\omega\sigma_b + \chi q^2 \qquad (2.4)$$

As in Eq. (1.7), we need the ratio of the imaginary and real parts of $D^{-1}$. We have

$$\frac{\text{Im}D^{-1}}{\text{Re}D^{-1}} \simeq \frac{\omega x t \tau_b}{\Delta(1-x)J\tau_f}, \quad q \leq q^* = \left[\frac{2\Delta\tau_f(1-x)J}{\chi}\right]^{1/2}$$

$$\simeq \frac{2\omega x t \tau_b}{\chi q^2}, \qquad q > q^*, \qquad (2.5)$$

where we have used Eqs. (1.5).

Therefore the recovered gauge free energy in the spin gap phase is:

$$F_s + N(0)\Delta^2/2 = \int_0^{\pi/a} \frac{d^2q}{(2\pi)^2} \int_0^{2\Delta} d\omega \coth\left(\frac{\omega}{2T}\right) \arctan\left[\frac{\text{Im}D^{-1}(q,\omega,\Delta)}{\text{Re}D^{-1}(q,\omega,\Delta)}\right], \qquad (2.6)$$

where $-N(0)\Delta^2/2$ is the contribution to the free energy of the spinon condensation. The susceptibility is still dominated by the fermion contribution, $\chi \simeq (1-x)J/(3k_f^2)$. For an example, we take $x = .2$ and find, in the limit $\Delta >> T$:

$$F_s = -3.06 N(0)\Delta^2 - .5 N(0)\Delta^2. \qquad (2.7)$$

Finally, we total the free energy cost (for $x = .2$) to condense the spinons:

$$F_s - F_n \simeq -35.5 N(0)\Delta^2[.100 - (J/\Delta)^{1/3}] \qquad (2.8)$$

Thus, in the dirty limit, the mean field spin gap can only exist if $\Delta >> J$. It is straightforward to check that the transition can only be first order.

### III. GAPLESS CASE

Gapless superconductivity can arise in a variety of ways. In the context of s-wave superconductivity, scattering from magnetic impurities causes breaking of singlet pairs and this leads to gapless superconductivity.[11,14] When $\Delta\tau_s < 1$ ($\tau_s$ is the spin-flip scattering rate), the density of states is non-zero at the Fermi level in the superconducting state and the dc conductivity has a finite value at $q = 0$. The superfluid density is reduced and the electron polarisability contains a conductivity-damping term $i\omega\sigma(q,\omega)$ down to zero frequency. When $\Delta\tau_s < 1$, $\sigma(q,\omega)$ is non-zero at $q = 0$, and grows linearly with frequency.

We shall assume that some spin-flip mechanism is at work in the strongly correlated medium, and that the spin-flip rate for the spinons is such that the condition $\Delta\tau_s < 1$ is satisfied. Here $\Delta$ is the fermion spin gap. Then the spin-gap phase behaves as a gapless superfluid. For simplicity, we take $\Delta\tau_s \simeq 1$. The conductivity in this case[14] can be approximated by $(\omega\sigma_n/6\Delta)$ for $\omega < 6\Delta$ and by $\sigma_n$ otherwise. With this choice of $\sigma$, we obtain the superfluid density by applying the $f$-sum rule: $\pi n/2m = \int \sigma(\omega)d\omega$, so that

$$\pi n_s/2m = \int (\sigma_{1n} - \sigma_{1s}) d\omega \qquad (3.1)$$

$$= \frac{\sigma_n}{6\Delta} \int_0^{6\Delta} \omega d\omega.$$

Therefore, in the spin gap phase, we have

$$\frac{n_s}{m_f} = \frac{6 n_f \Delta \tau_s}{\pi m_f} \qquad (3.2)$$

(this is strictly valid for $\Delta \tau_s < \pi/6$) and

$$\sigma_s = \frac{\omega \sigma_n}{6\Delta} = \frac{\pi \omega n_s/m_f}{36 \Delta^2}, \quad \omega < 6\Delta \qquad (3.3)$$

As usual, in order to assess the stability of the spin gap phase, we compare its energy with that of the normal phase. We look at the cost in the gauge field contribution to the free energy which is paid upon entry into the spinon condensate phase. It is again given by $F_{s,gauge} - F_{n,gauge}$, where $s$, $n$ stand for the spinon condensate and normal phases. Usually, $F_{s,gauge} = F_{n,gauge}$ for frequencies larger than $2\Delta$. In the gapless situation, we have seen that the spinon conductivity in the condensed phase is only equal to that in the normal phase for frequencies above $6\Delta$. Then the gauge field free energy differs between the normal and condensed phases up to a frequency $6\Delta$, after which the conductivities (hence ImΠ) in the two phases are equal. Therefore, we have to evaluate the free energy contributions up to frequency $6\Delta$. In the normal phase, ImΠ $\simeq \omega \sigma_f = \omega (n_f/m_f) \tau_s$ and ReΠ $\simeq \chi_f q^2 = (q^2/12\pi m_f)$. The spin-flip scattering rate is not reduced by screening effects. In the normal state, we have a contribution which differs from that of the previous sections since the absence of screening makes the fermion conductivity (now given by $n_f \tau_s/m_f$) exceed the Landau damping [given by $(n_f/m_f v_f q)$] over a larger portion of the Brillouin zone. For simplicity we take the mean free path of order the lattice spacing, implying that the spin-scattering conductivity determines ImΠ over the whole zone. This is an extreme case which favors the spin gap phase. Consequently, we write, in the normal phase,

$$\frac{\text{Im}\Pi}{\text{Re}\Pi} = \frac{\omega \tau_s n_f/m_f}{q^2/12\pi m_f}. \qquad (3.4)$$

To determine the free energy loss from the normal phase, we use this expression in the RHS of Eq. (1.7) and integrate the frequency to $6\Delta$. The result is (for $x = .2$ and $\Delta \tau_s = 1$)

$$F_n = -271 N(0) \Delta^2 \frac{J}{\Delta}. \qquad (3.5)$$

We again recover a gauge field contribution in the condensed phase since the conductivity does not vanish there. The calculation of the gauge field free energy in the spinon condensed phase is similar to that of Sec. IIB. In the present case however, the fermion conductivity [Eq. (3.3)] is not zero for $\omega < 6\Delta$. In fact, for $x = .2$, it is larger than the boson conductivity which was used in Eq. (2.3) as soon as $\omega \gtrsim .004\Delta$, so we use

$$\text{Im}\Pi = -\omega \sigma_s \simeq -\frac{n_f \omega^2 \tau_s}{m_f 6\Delta} \qquad (3.6)$$

for all $\omega$ up to $6\Delta$.

At the same time the superfluid stiffness adds a constant part to ReΠ:,

$$\text{Re}\Pi \simeq \frac{6}{\pi} \left( \frac{n_f}{m_f} \Delta \tau_s \right) + \chi q^2. \qquad (3.7)$$

This first term on the RHS exceeds the second over the whole zone. Consequently, in the spinon condensate phase, we have

$$\frac{\text{Im}\Pi}{\text{Re}\Pi} = \pi \left( \frac{\omega}{6\Delta} \right)^2. \qquad (3.8)$$

We use this in Eq. (2.5) and replace the upper limit of the $\omega$ integration by $6\Delta$. The result is

$$F_{s,gauge} = -95.6 N(0) \Delta^2 (J/\Delta) \qquad (3.9)$$

The condensation energy is much reduced in the gapless phase compared to the pure case; with our choice $\Delta \tau_{sf} \simeq 1$, it is reduced to:[14]

$$F_{cond} = -.05 N(0) \Delta^2 \tag{3.10}$$

The net free energy cost is

$$F_{s,gauge} + F_{cond} - F_{n,gauge} = -175 N(0) \Delta^2 [.0003 - (J/\Delta)]. \tag{3.11}$$

Thus, we see once again that the cost in zero point energy exceeds by far the gain in condensation energy since the latter is so small in the gapless case.

## IV. CONCLUSION

The mean field spin gap is destroyed by the gauge fluctuations, in all the cases considered so far: pure case,[9] dirty limit, gapless phase (this work), in the absence of singular interactions. We have not considered the case of an anisotropic spinon condensate, say "$d$-wave". In the case of a condensate with nodes, there are again gapless excitations and a reduction in condensation energy. However, for the case of point nodes, as in the two-dimensional $d$-wave case, the condensation energy remains of order $\Delta^2$. Furthermore, as Übbens and Lee[9] have shown, the presence of a non-zero spinon density of states and the fact that for some momentum transfers the gauge field-spinon interaction is pair-enhancing rather than pair-breaking does not change their conclusion. Thus, we do not expect a result which differs from that of Section III. It might be argued that since impurity scattering is pairbreaking for a gap function with nodes, it would effectively destroy a $d$-wave spinon condensate anyway. However this effect arises when the scattering connects two regions of the Fermi surface which have gap functions of opposite sign. However, as discussed in Section IIA, the spinon-impurity scattering is predominately of small momentum transfer and therefore most of the scattering connects gap regions of the same sign.

We have not discussed the contribution of longitudinal gauge fluctuations. This might seem dangerous in view of the findings of Ref. 10 according to which longitudinal fluctuations have a dramatic effect at high temperature on the total entropy of the spinon-holon system; they reduce this entropy by roughly a factor two at low doping. However, at low temperature, longitudinal fluctuations are frozen, as in an ordinary charged electron liquid, so that they have exponentially small contributions in both the spin gap phase and the strange metal phase. Therefore we do not think they change significantly the picture obtained when considering transverse fluctuations alone.

It is not clear whether the disappearance of the spin gap phase in the gauge theory of the single-layer material improves the agreement between that theory and experiment. Übbens and Lee[9] adopt the point of view of Millis and Monien,[15] who argued, on the basis of NMR and neutron diffraction data, that the spin gap phase exists in the two-layer compound YBCO (yttrium barium copper oxide) but is absent in the single-layer LSCO (lanthanum cuprate). Then they proceed to show that an interlayer pairing mechanism in YBCO restores the spin gap in that compound. However, the evidence against the spin gap in LSCO is not completely unambiguous. It has been argued that similar suppressions in magnetic, transport and thermodynamic properties is rather generic in *all* underdoped materials[16,17]. It is tempting to ascribe these observations to a spingap mechanism. However, according to the results of Übbens and Lee, as well as of this paper, the slave boson gauge theory cannot account for this suppression of inelastic scattering processes in terms of suppressed spin degrees of freedom in the single layer materials.

While there is certainly disagreement at present, if it turns out that both LSCO and YBCO have quite similar "spin-gap" phenomenology, it will be difficult to account for the behavior by different mechanisms for the two compounds. Then the inability of the present slave-boson gauge-field approach to describe a spin gap in the normal phase of LSCO will be a serious shortcoming of that approach and raise questions about its use to analyze normal state properties of YBCO.[5,18,19]

In a recent work, Wen and Lee[20] have introduced a new mean-field theory for the single-layer $t - J$ model. There is a staggered flux phase which has a spin gap. Since the gauge field in the approach remains massless in the spin gap phase, the latter may turn out to be stable even after consideration of the fluctuations.

## V. ACKNOWLEDGEMENTS

We acknowledge helpful discussions with J.P. Rodriguez, L.B. Ioffe, H. Fukuyama, and J. Sauls. This work was supported in part by CEE contract Capital Humain et Mobilité CHRX-CT93-00332 (PL), by NSF Grant 92-21907 (EA) and by the Centre International des Etudiants Stagiaires (EA). EA acknowledges the hospitality of both the Groupe de Physique des Solides at Orsay and the Aspen Center for Physics.